\documentclass[3p,times]{elsarticle}
\usepackage{subfigure}
\usepackage{tikz}
\usetikzlibrary{trees,arrows}
\usepackage{float} % note: erase [H] to let latex locate it
\usepackage{url}
\usepackage{graphicx}
\usepackage{amssymb}
\usepackage{amsmath}
\usepackage{epstopdf}
\usepackage{hyperref}
\usepackage{pdflscape}
\usepackage{xcolor}
\usepackage[font=small]{caption}

\newcommand\var{{\operatorname{\mathbf{var}}}}

\usepackage{xcolor}
\newcommand\ytl[2]{
\parbox[b]{8em}{\hfill{\color{gray}\bfseries\sffamily #1}~$\cdots$~}\makebox[0pt][c]{$\bullet$}\vrule\quad \parbox[c]{9.5cm}{\vspace{7pt}\color{darkgray}\raggedright\sffamily #2.\\[7pt]}\\[-3pt]}

\begin{document}
\begin{frontmatter}
%\newcolumntype{C}{>{\centering\arraybackslash}p{5em}}
%\newcolumntype{D}{>{\centering\arraybackslash}p{8em}}

\title{Zero  %variation of the 
Black-Derman-Toy interest rate model}

\author[label1]{Grzegorz Krzy\.zanowski}
\ead{grzegorz.krzyzanowski@pwr.edu.pl}

\author[label2]{Ernesto Mordecki}
\ead{mordecki@cmat.edu.uy}

\author[label3]{Andr\'es Sosa}
\ead{asosa@iesta.edu.uy}

\address[label1]{Hugo Steinhaus Center,
Faculty of Pure and Applied Mathematics, Wroclaw University of Science and Technology
50-370 Wroclaw, Poland}

\address[label2]{Centro de Matemática, Facultad de Ciencias, Universidad de la República}

\address[label3]{Instituto de Estad\'istica - Facultad de Ciencias Económicas y de Administraci\'on - Universidad de la República, Uruguay}

%%Grzegorz Krzy{\.{z}}anowski\footnote{Faculty of Pure and Applied Mathematics, Wroc{\l}aw University of Science and Technology. email: grzegorz.krzyzanowski@pwr.edu.pl}, Ernesto Mordecki\footnote{Centro de Matemática, Facultad de Ciencias, Universidad de la República. email: mordecki@cmat.edu.uy}, Andr\'es Sosa\footnote{Centro de Matemática, Facultad de Ciencias, Universidad de la República. email: asosa@cmat.edu.uy}
%}
%\maketitle

\begin{abstract}
We propose a modification of the classical Black-Derman-Toy (BDT)
interest rate tree model, 
which includes the possibility of a jump with small probability at each step
to a practically zero interest rate. 
The corresponding BDT algorithms are consequently modified to calibrate the tree
containing the zero interest rate scenarios. 
This modification is motivated by the recent 2008--2009 crisis in the  United States 
and it quantifies the risk of a future crises in bond prices and derivatives.
The proposed model is useful to price derivatives. This exercise also provides a tool to 
calibrate the probability of this event. A comparison of option prices and implied volatilities on US Treasury bonds
computed with both the proposed and the classical tree model is provided, in six different scenarios along the different periods comprising the years 2002--2017.
\end{abstract}

%\newpage
%\tableofcontents
%\newpage

\end{frontmatter}
\section{Introduction}

The Federal Funds Rate (i.e., the interest rate at which depository institutions lend reserve balances to other depository institutions overnight
on an uncollateralized basis) is an important benchmark in financial markets. This interest rate affects monetary and financial conditions  which influence certain aspects of the general economy  in the United States, such as employment, growth, inflation and term structure interest rates.

\begin{figure}%[H]
\centering 
\includegraphics[width=13.5cm,height=9.0cm]{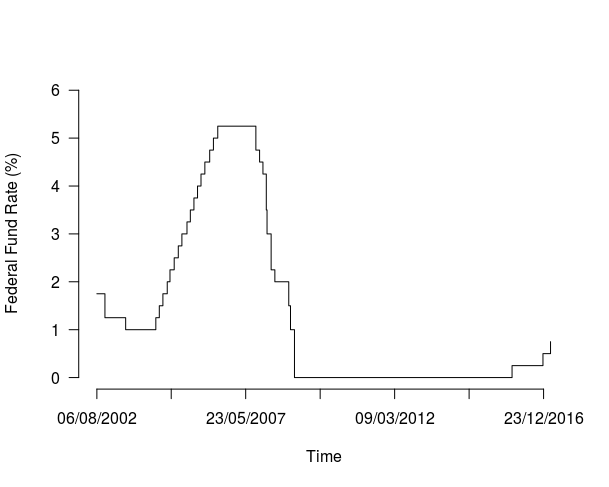}
\caption{Federal Fund Rate (2002-2017).}\label{figure:fed}
\end{figure}

Following the 2007--2008 financial crisis in United States,  the Federal Reserve reduced the Fed Funds Rate by 425 basis points to
practically  zero (targeting interest rates in the interval $0$-$0.25\%$) in one year.  This decision was preserved for nine years and was called
the Zero Interest Rate Policy  (ZIRP policy). Figure \ref{figure:fed} shows the evolution of the Federal Funds Rate between the years 2002 and 2017.

Motivated by this phenomena, 
and inspired by  Lewis's (2016) 
ZIRP models  in  continuous time, 
and by using the default models of Duffie and Singleton (1999),
we propose a modification of the classical Black-Derman-Toy  (BDT) model on interest rates (Black, Derman \&  Toy, 1990).
\subsection{Different approaches to  model  the ZIRP}
Recently, several approaches to model  the ZIRP have appeared. 
Lewis (2016) makes two proposals, which he summarizes as: 
(i) slowly-reflecting boundaries, also known as sticky boundaries; and 
(ii) jump-returns from a boundary.
The first model consists in the utilization of a resource used in diffusions considered as Markovian processes,
consisting of the introduction of sticky points. 
The sticky point retains the process for a longer time than the other points. 
To produce this phenomena, in the continuous time model, 
an atomic point is introduced in the speed measure of the diffusion (Borodin \&  Salminen, 2002).
The second model consists of the introduction of a delayed start of the process. 
This delay time is modeled by an exponential random variable.
The process stays at the $x=0$ level until this exponential time. 
It then jumps to an independent state, from which it continues its dynamics as a diffusion.
The bond prices for these models are given in  (Lewis, 2016).

%\subsubsection{Skew CIR (chinese)\textcolor{red}{Skew CIR process}}

An alternative approach was proposed by Tian and Zhang (2018).
These authors depart from the classical CIR process (Cox, Ingresoll \&  Ross, 1985), 
and add one skew point at a certain relatively small level of the interest rate.
The skew phenomena in diffusion models represents a permeable barrier.
When the process reaches the skew point, the probability of upwards and downwards movements is modified
according to a certain probability.  
In this way, if the probability of downwards continuation is higher that $1/2$, as the CIR process never reaches zero,
then the proposed process remains below the skew point for a longer time than the CIR process.
The skew diffusions can be constructed by departing from the excursion theory for diffusions,
and in many other ways (Lejay, 2006). The discrete analogue of this model
is a binary random walk with symmetric probabilities at all states with the exception of one---the skew point.
At this point there is a higher probability of going downwards. 
This produces a process that stays longer below
the critical threshold than the original. 
It also can be seen that the weak limit of this process, properly normalized,
goes to a skew diffusion (Lejay, 2006). 
In the paper (Tian \&  Zhang, 2018), 
based on stochastic calculus arguments, 
the authors give bond prices for this model.

Another approach to model the ZIRP phenomena was introduced by  
Eberlein et al.(2018).
This proposal is in the context of L\'evy modeling of Libor rates, 
and the modification allows negative interest rates.
This model is especially suited for calibration in the presence of extremely low rates,
it is presented in the framework of the semimartingale theory, 
and includes derivatives pricing,  particularly caplets. 
As an application, European caplets market prices are used to calibrate the proposed model,
with the help of Normal inverse Gaussian L\'evy processes.

%\subsubsection{An Overview of Post-crisis Term Structure Models}

Martin (2018) made an alternative proposal, which 
considers that the financial crisis changes the modeling perspective of the term structure. 
The main reason is that there are  differences between interest rates that were previously linked. 
Therefore, 
the proposal is to use several interest rate curves in the same model, 
which reflect the different types of risk observed in the fixed income markets.
The paradigm of the valuation that the authors use is based on intensity models. 
The dynamics of the term structure is given by exponential affine factor models.  
The hazard rate  incorporates the risk observed in the interbank sector that affects the corresponding interest rate.
The author states that the approach is important for long-term assets, such as swaps and swaptions.

%\subsubsection{Explicit Computation of the Post-crisis Spot LIBOR in a Jump-Diffusion Framework}

%Let us finally mention the  paper by Di Persio and Gugole,
%\cite{DiPersioGugole}, where the authors use the intensity approach to include jumps in the Libor rate stochastic
%evolution to model the observed 
%significant spreads between the same interbank rate, e.g. the LIBOR rate, considered at different tenors.

\subsection{Our proposal}

In view of the need of adequate models to the ZIRP, 
we propose to depart from the Black-Derman-Toy (BDT) binary tree model, 
incorporating into its dynamics
the possibility of a downwards jump with a small probability at each time step
to a practically zero interest rate value.
Additionally, we assume that
once the process reaches the zero interest rate zone,
it remains there with high probability.
This proposal mimics the intensity approach in default bond models proposed by Duffie and Singleton (1999),
by jumping to near zero according to a geometric random variable with a small rate. 
In addition, the sticky phenomena described by Lewis (2016), 
as the interest rate process, 
once this jump is realized, 
stays with high probability in this close to zero zone.
In practical terms, 
the initial BDT binary tree model
is modified to a mixed binary-ternary tree model 
to find consistent interest rates with the market term structure. 
The new model is called the ZBDT model (Zero  Black-Derman-Toy interest rate model).

%\subsection{Detailed Contents}
The rest of this paper is structured as follows.
In Section \ref{section:bdt}, we introduce the main ideas of the classical BDT model with an emphasis on calibration,
with the aim of introducing  the ZBDT model in Section \ref{section:zbdt}, together
with its respective calibration equations.
Section \ref{section:empirical} has an empirical content. 
It  contains a detailed account of critical financial events during the period of study (2002--2017) in the United States,
which provides information about interest rates with their respective volatilities, and we choose
six representative different scenarios to compare the results given by the BDT and the ZBDT models.
In Section \ref{section:conclusions}, we conclude with a brief discussion of the results and comments on some possible future work.

\section{The Black-Derman-Toy  model}\label{section:bdt}

%\subsection{Motivation}

The Black-Derman-Toy model (Black et al. 1990) is one of the most popular and celebrated models in fixed income interest rate theory. 
It consists in a binary tree with equiprobable transitions, which makes it simple and flexible to use.
More precisely, the model departs from the current interest rate curve, from where the yields for different maturities are extracted, and it uses a series of consecutive historical interest rate curves during a certain time interval to compute this yield volatilities.
The model assumes that the volatility only depends on time and not on the value of the interest rate.
A calibration procedure is implemented to obtain the interest rates acting during the respective
time intervals defined in the model. 

The model assumes that the future interest rates evolve randomly in a binomial tree 
with two scenarios at each node, labeled, respectively, by $u$ (for ``up") and $d$ (for ``down"), 
with the particularity that an $u$ followed by a $d$ take us to the same value as a $d$ followed by an $u$.
In this way, after $n$ periods, we have $n+1$ possible states for our stochastic process modeling the interest rate.
With the aim of simplifying the presentation, we consider that one period is equivalent to one year. Moreover, in whole paper we focus on the zero-coupon  bonds (zc-bonds). The corresponding modification
to shorter periods or use the bonds with coupons is straightforward.
In Figure \ref{figure:bdtp} (b), we present the tree corresponding to the prices of a zc-bond 
with expiration in $n=3$ years,
where we denote by $B_{ij}$ the zc-bond price corresponding to the period $i$ and state $j$
for the same values of $i$ and $j$.
Here and in whole paper we assume that the face value (FV) of the bond equals $100$: $B_{nj}=100$ for $j=1,\dots,n+1$.

\begin{figure}%[H]
\centering
\subfigure[]{
\begin{tikzpicture}[scale=0.82, transform shape]
\tikzstyle{every node} = [circle, fill=gray!0,draw]
\node (11) at (0,0) {$r_{0,1}$};
\node (21) at (3,0) {$r_{1,1}$};
\node (22) at (3,2) {$r_{1,2}$};
\node (31) at (6,0) {$r_{2,1}$};
\node (32) at (6,2) {$r_{2,2}$};
\node (33) at (6,4) {$r_{2,3}$};
\tikzstyle{every node} = [circle, fill=gray!40]
\node (1121) at (1.5,0) {$\frac12$};
\node (1122) at (1.5,1) {$\frac12$};
\node (2131) at (4.5,0) {$\frac12$};
\node (2132) at (4.5,1) {$\frac12$};
\node (2232) at (4.5,2) {$\frac12$};
\node (2233) at (4.5,3) {$\frac12$};
\foreach \from/\to in {11/1121,11/1122}
\draw [-,line width=0.4mm,fill=black] (\from) -- (\to);
\foreach \from/\to in {1121/21,1122/22}
\draw [->,line width=0.4mm,fill=black] (\from) -- (\to);
\foreach \from/\to in {21/2131,21/2132,22/2232,22/2233}
\draw [-,line width=0.4mm,fill=black] (\from) -- (\to);
\foreach \from/\to in {2131/31,2132/32,2232/32,2233/33}
\draw [->,line width=0.4mm,fill=black] (\from) -- (\to);

\end{tikzpicture}
}\hspace{15mm}
\centering
\subfigure[]{
\begin{tikzpicture}[scale=0.82, transform shape]
\tikzstyle{every node} = [circle, fill=gray!0,draw]
\node (11) at (0,0) {$B_{0,1}$};
\node (21) at (3,0) {$B_{1,1}$};
\node (22) at (3,2) {$B_{1,2}$};
\node (31) at (6,0) {$B_{2,1}$};
\node (32) at (6,2) {$B_{2,2}$};
\node (33) at (6,4) {$B_{2,3}$};
\node (41) at (9,0) {$100$};
\node (42) at (9,2) {$100$};
\node (43) at (9,4) {$100$};
\node (44) at (9,6) {$100$}; 
\tikzstyle{every node} = [circle, fill=gray!40]
\node (1121) at (1.5,0) {$r_{0,1}$};
\node (1122) at (1.5,1) {$r_{0,1}$};
\node (2131) at (4.5,0) {$r_{1,1}$};
\node (2132) at (4.5,1) {$r_{1,1}$};
\node (2232) at (4.5,2) {$r_{1,2}$};
\node (2233) at (4.5,3) {$r_{1,2}$};
%%%%%%%%%%%%%%%%%%%
\node (3141) at (7.5,0) {$r_{2,1}$};
\node (3142) at (7.5,1) {$r_{2,1}$};
\node (3242) at (7.5,2) {$r_{2,2}$};
\node (3243) at (7.5,3) {$r_{2,2}$};
\node (3343) at (7.5,4) {$r_{2,3}$};
\node (3344) at (7.5,5) {$r_{2,3}$};
\foreach \from/\to in {11/1121,11/1122}
\draw [-,line width=0.4mm,fill=black] (\from) -- (\to);
\foreach \from/\to in {1121/21,1122/22}
\draw [->,line width=0.4mm,fill=black] (\from) -- (\to);
\foreach \from/\to in {21/2131,21/2132,22/2232,22/2233}
\draw [-,line width=0.4mm,fill=black] (\from) -- (\to);
\foreach \from/\to in {2131/31,2132/32,2232/32,2233/33}
\draw [->,line width=0.4mm,fill=black] (\from) -- (\to);
\foreach \from/\to in {31/3141,31/3142,32/3242,32/3243,33/3343,33/3344}
\draw [-,line width=0.4mm,fill=black] (\from) -- (\to);
\foreach \from/\to in {3141/41,3142/42,3242/42,3243/43,3343/43,3344/44}
\draw [->,line width=0.4mm,fill=black] (\from) -- (\to);
\end{tikzpicture}
}

\caption{The BDT interest rate tree (a) and the corresponding tree of zc-bond (b) with $T=3$ and $FV=100$.}\label{figure:bdtp}
\end{figure}

%\begin{figure}%[H]
%\centering
%\begin{tikzpicture}[scale=0.7, transform shape]
%\tikzstyle{every node} = [circle, fill=gray!0,draw]
%\node (11) at (0,0) {$r_{0,1}$};
%\node (21) at (3,0) {$r_{1,1}$};
%\node (22) at (3,2) {$r_{1,2}$};
%\node (31) at (6,0) {$r_{2,1}$};
%\node (32) at (6,2) {$r_{2,2}$};
%\node (33) at (6,4) {$r_{2,3}$};
%\foreach \from/\to in {11/21,11/22}
%\draw [->,line width=0.4mm,fill=black] (\from) -- (\to);
%\foreach \from/\to in {21/31,21/32,22/32,22/33}
%\draw [->,line width=0.4mm,fill=black] (\from) -- (\to);
%\end{tikzpicture}
%
%\end{figure}

The evolution of this bond is associated to a tree with the interest rates that 
apply to each time period, as shown in Figure \ref{figure:bdtp} (a).
In the BDT model the probability of each $u$ or $d$ scenario at each node is $1/2$, the evolutions are independent, 
and the values of the interest rates are obtained through calibration.

%This fact should be contrasted to the Cox-Ross-Rubinstein binomial tree for floating rate derivatives  \cite{crr}
%where the financial security modeled exhibits the same dynamic at all different nodes.
%This dynamics  that depends on the only one parameter (the volatility), that gives the up and down movements
%and also the respective probabilities.

\subsection{Calibration of the BDT model}

In a model with $n$ time periods, 
we calibrate a tree of order $n$ departing from the following data:
the yields  on zc-bonds $y(k),\ k=1,\dots,n$, corresponding to the respective periods $[0,k]$ (the first $k$ periods),
and the yield volatilities for the same bonds $\beta(k),\ k=2,\dots,n$, under the same convention.

The interest rates of the tree, are $\{r_{i,j}\colon i=0,\dots,n-1; j=1,\dots,i+1\}$, 
and correspond to each time period in the up and down scenarios, 
giving $n(n+1)/2$ unknowns to be calibrated.

The first step uses only $y(1)$ and concludes that $r_{0,1}=y(1)$:
\begin{align*}
B_{1,1}&=B_{1,2}=100,\\
B_{0,1}&=\frac{100}{1+y(1)}=\frac12\,\frac{1}{1+r_{0,1}}\left(B_{1,1}+B_{1,2}\right).
\end{align*}
When $n>1$, we introduce the yields $y_u$ (up) and $y_d$ (down) one year from now,
corresponding to bond prices $B_u$ and $B_d$. The relevant relations that this quantities 
satisfy are
\begin{equation*}
B_u=\frac1{(1+y_u)^{n-1}},\quad B_d=\frac1{(1+y_d)^{n-1}}.
\end{equation*}

\subsubsection*{Variance equation at a node}

Consider a tree with $n$ steps.
We introduce a random variable $Y$ that takes two values:
$$
Y=
\begin{cases}
y_u,&\text{ with probability $1/2$},\\
y_d, &\text{ with probability $1/2$}.
\end{cases}
$$
Then, $\log Y$ has a variance 
$
\var\log Y=\beta^2(n),\text{ if and only if } y_u=y_de^{2\beta(n)},
$
equivalent to
\begin{equation}
\label{eq:var}
\beta(n)=\frac12\log{y_u\over y_d},
\end{equation}
as follows from the following computation:
\begin{multline*}
\var\log Y=\frac{1}{2}\log^2y_u+\frac{1}{2}\log^2y_d-\left(\frac{1}{2}(\log y_u+\log y_d)\right)^2
	=\left(\frac{1}{2}(\log y_u -\log y_d)\right)^2=\left(\frac{1}{2}\log\frac{y_u}{y_d}\right)^2
	%\\
	%=\left(\frac{1}{2}\log(\frac{r_{i+1,j}\exp(2\sigma(i+1))}{r_{i+1,j}})\right)^2
	=\beta(n)^2.
\end{multline*}
%$$
%\log {y\over y_0}=
%\begin{cases}
%\log {y_u \over y_0},&\text{ with probability}\   \hat{p},\\
%\log {y_d \over y_0}, &\text{ with probability}\  \hat{p},\\
%0, &\text{ with probability}\  p.\\
%\end{cases}
%$$
%The mean of $\log(y/y_0)$ is
%$$
%\frac{1-p}2\left(\log{y_u\over y_0}+\log{y_d\over y_0}\right),
%$$
%then
%$$
%\var\log {y\over y_0} =\frac{1-p}2\left(\left(\log{y_u\over y_0}\right)^2+\left(\log{y_d\over y_0}\right)^2\right)-
%\left(
%\frac{1-p}2\left(\log{y_u\over y_0}+\log{y_d\over y_0}\right)
%\right)^2.
%$$
%This is why we introduce
%\begin{equation}\label{eq:ells}
%\ell_u=\log\frac{y_{u}}{y_{0}},\quad
%\ell_d=\log\frac{y_{d}}{y_{0}}.
%\end{equation}
%to obtain
%$$
%\aligned
%\var\log y&=\frac{1-p}2\left(\ell_u^2+\ell_d^2\right)-
%\left(
%\frac{1-p}2\left(\ell_u+\ell_d\right)
%\right)^2
%\\
%&=\frac{1-p^2}4\left(\ell_u^2+\ell_d^2\right)-\frac{(1-p)^2}2\ell_u\ell_d.
%\\
%%&=\frac{1-p^2}4\left(\ell_u-\ell_d\right)^2-p(1-p)\ell_u\ell_d.
%\endaligned
%$$
%
%
%
The BDT model assumes that the variance of the log-interest rate  
 with fixed time
is constant for all nodes.
The respective interest rates at each node at time $n-1$ are  
represented by an auxiliar random variable $R_{n-2,j}$. 
$$
R_{n-2,j}=
\begin{cases}
r_{n-1,j+1},&\text{ with probability $1/2$},\\
r_{n-1,j}, &\text{ with probability $1/2$},\\
\end{cases}
$$
for $j=1,\dots,n-1$.
The variance of this random variable is assumed to be constant for all nodes at the same time period, and
satisfies
\begin{equation}\label{eq:sigma}
\sigma(n)=\frac12\log{r_{n-1,j+1}\over r_{n-1,j}},\quad j=1,\dots,n-1.
\end{equation}
In the second step of the calibration, the new data are $y(2)$ and $\beta(2)$.
The unknowns are $r_{1,1},r_{1,2}$ and $\sigma(2)$.
In this case $\sigma(2)=\beta(2)$,
because the local variation of the interest rate for one year coincides with the global variation.
Accordingly, $y_u=r_{1,2}$ and $y_d=r_{1,1}$.
The bond prices then satisfy
\begin{align*}
B_{0,1}&=\frac{100}{{(1+y(2))}^2}=\frac12\frac{1}{(1+r_{0,1})}\left(B_{u}+B_{d}\right),\\
B_{2,j}&=100, j=1,2,3,\\
B_{u}&=\frac{100}{{(1+y_u)}},\quad B_{d}=\frac{100}{{(1+y_d)}},\\
\beta(2)&=\frac12\log{y_u\over y_d}.\\
\end{align*}
For general $n$ the new data are $y(n)$ and $\beta(n)$. The unknowns are $r_{n-1,j}$ for $j=1\dots,n$
and $\sigma(n)$.
The value $\sigma(n)^2$ is the variance of the interest rate at each node (see \eqref{eq:var}).
The bond prices then satisfy

\begin{align*}
\frac{100}{{(1+y(n))}^{n}}&=\frac12\frac{1}{(1+r_{0,1})}\left(B_{u}+B_{d}\right),\\
B_{n,j}&=100, \  j=1,\dots,(n+1),\\
B_{i,j}&=\frac12\frac{1}{(1+r_{i,j})}\left({B_{i+1,j}+B_{i+1,j+1}}\right),\  i=0,\dots,(n-1), \ j=1,\dots,i+1,\\
B_{u}&=\frac{100}{{(1+y_u)}^{n-1}},\quad B_{d}=\frac{100}{{(1+y_d)}^{n-1}},\\
\beta(n)&=\frac12\log{y_u\over y_d},\\
\sigma(n)&=\frac12\log{r_{n-1,j}\over r_{n-1,j-1}}, \ \ j=2,\dots,n.
\end{align*}

\section{The ZBDT model}\label{section:zbdt}

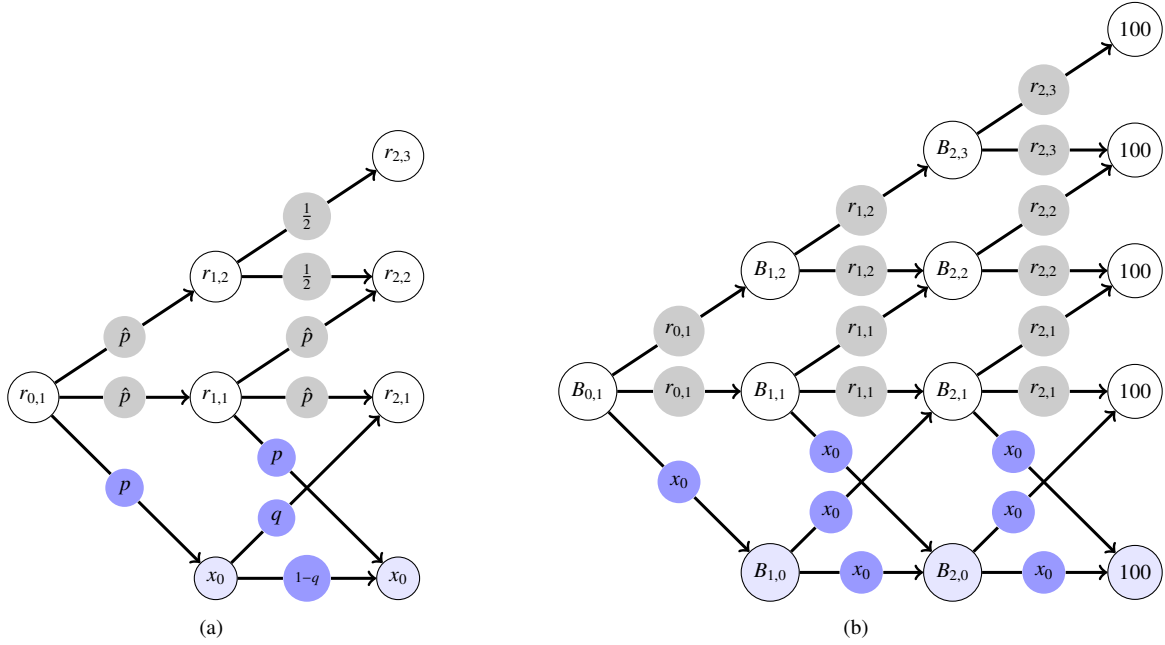
\begin{figure}%[H]
\centering
\subfigure[]{
\begin{tikzpicture}[scale=0.8, transform shape]
\tikzstyle{every node} = [circle, fill=gray!0,draw]
\node (11) at (0,0) {$r_{0,1}$};
\node (21) at (3,0) {$r_{1,1}$};
\node (22) at (3,2) {$r_{1,2}$};
\node (31) at (6,0) {$r_{2,1}$};
\node (32) at (6,2) {$r_{2,2}$};
\node (33) at (6,4) {$r_{2,3}$};
%\node (41) at (9,0) {$100$};\hat{p}
%\node (42) at (9,2) {$100$};
%\node (43) at (9,4) {$100$};
%\node (44) at (9,6) {$100$}; 
\tikzstyle{every node} = [circle, fill=gray!40]
\node (1121) at (1.5,0) {$\hat{p}$};
\node (1122) at (1.5,1) {$\hat{p}$};
\node (2131) at (4.5,0) {$\hat{p}$};
\node (2132) at (4.5,1) {$\hat{p}$};
\node (2232) at (4.5,2) {$\frac12$};
\node (2233) at (4.5,3) {$\frac12$};
%%%%%%%%%%%%%%%%%%%
%\node (3141) at (7.5,0) {$\hat{p}$};
%\node (3142) at (7.5,1) {$\hat{p}$};
%\node (3242) at (7.5,2) {$\frac12$};
%\node (3243) at (7.5,3) {$\frac12$};
%\node (3343) at (7.5,4) {$\frac12$};
%\node (3344) at (7.5,5) {$\frac12$};
\foreach \from/\to in {11/1121,11/1122}
\draw [-,line width=0.4mm,fill=black] (\from) -- (\to);
\foreach \from/\to in {1121/21,1122/22}
\draw [->,line width=0.4mm,fill=black] (\from) -- (\to);
\foreach \from/\to in {21/2131,21/2132,22/2232,22/2233}
\draw [-,line width=0.4mm,fill=black] (\from) -- (\to);
\foreach \from/\to in {2131/31,2132/32,2232/32,2233/33}
\draw [->,line width=0.4mm,fill=black] (\from) -- (\to);
%\foreach \from/\to in {31/3141,31/3142,32/3242,32/3243,33/3343,33/3344}
%\draw [-,line width=0.4mm,fill=black] (\from) -- (\to);
%\foreach \from/\to in {3141/41,3142/42,3242/42,3243/43,3343/43,3344/44}
%\draw [->,line width=0.4mm,fill=black] (\from) -- (\to);
%%% ZBDT
\tikzstyle{every node} = [circle, fill=blue!10,draw]
\node (20) at (3,-3) {$x_0$};
\node (30) at (6,-3) {$x_0$};
%\node (40) at (9,-3) {$100$};
\tikzstyle{every node} = [circle, fill=blue!40]
\node (1120) at (1.5,-1.5) {$p$};
\node (2030) at (4.5,-3) {$\scriptstyle1-q$};
%\node (3040) at (7.5,-3) {$\scriptstyle1-q$};
%%%%%%%%% 
\node (2130) at (4,-1) {$p$};
\node (2031) at (4,-2) {$q$};
%\node (3140) at (7,-1) {$p$};
%\node (3041) at (7,-2) {$q$};
%%%%%%%%% 
\foreach \from/\to in {11/1120}
\draw [-,line width=0.4mm,fill=black] (\from) -- (\to);
\foreach \from/\to in {1120/20}
\draw [->,line width=0.4mm,fill=black] (\from) -- (\to);
\foreach \from/\to in {20/2030}
\draw [-,line width=0.4mm,fill=black] (\from) -- (\to);
\foreach \from/\to in {2030/30}
\draw [->,line width=0.4mm,fill=black] (\from) -- (\to);
%\foreach \from/\to in {30/3040}
%\draw [-,line width=0.4mm,fill=black] (\from) -- (\to);
%\foreach \from/\to in {3040/40}
%\draw [->,line width=0.4mm,fill=black] (\from) -- (\to);
%%%%%%%%% 
\foreach \from/\to in {20/2031,21/2130}%,30/3041,31/3140}
\draw [-,line width=0.4mm,fill=black] (\from) -- (\to);
\foreach \from/\to in {2130/30,2031/31}%,3140/40,3041/41}
\draw [->,line width=0.4mm,fill=black] (\from) -- (\to);
\end{tikzpicture}
}\hspace{15mm}
\centering
\subfigure[]{
\begin{tikzpicture}[scale=0.8, transform shape]
\tikzstyle{every node} = [circle, fill=gray!0,draw]
\node (11) at (0,0) {$B_{0,1}$};
\node (21) at (3,0) {$B_{1,1}$};
\node (22) at (3,2) {$B_{1,2}$};
\node (31) at (6,0) {$B_{2,1}$};
\node (32) at (6,2) {$B_{2,2}$};
\node (33) at (6,4) {$B_{2,3}$};
\node (41) at (9,0) {$100$};
\node (42) at (9,2) {$100$};
\node (43) at (9,4) {$100$};
\node (44) at (9,6) {$100$}; 
\tikzstyle{every node} = [circle, fill=gray!40]
\node (1121) at (1.5,0) {$r_{0,1}$};
\node (1122) at (1.5,1) {$r_{0,1}$};
\node (2131) at (4.5,0) {$r_{1,1}$};
\node (2132) at (4.5,1) {$r_{1,1}$};
\node (2232) at (4.5,2) {$r_{1,2}$};
\node (2233) at (4.5,3) {$r_{1,2}$};
%%%%%%%%%%%%%%%%%%%
\node (3141) at (7.5,0) {$r_{2,1}$};
\node (3142) at (7.5,1) {$r_{2,1}$};
\node (3242) at (7.5,2) {$r_{2,2}$};
\node (3243) at (7.5,3) {$r_{2,2}$};
\node (3343) at (7.5,4) {$r_{2,3}$};
\node (3344) at (7.5,5) {$r_{2,3}$};
\foreach \from/\to in {11/1121,11/1122}
\draw [-,line width=0.4mm,fill=black] (\from) -- (\to);
\foreach \from/\to in {1121/21,1122/22}
\draw [->,line width=0.4mm,fill=black] (\from) -- (\to);
\foreach \from/\to in {21/2131,21/2132,22/2232,22/2233}
\draw [-,line width=0.4mm,fill=black] (\from) -- (\to);
\foreach \from/\to in {2131/31,2132/32,2232/32,2233/33}
\draw [->,line width=0.4mm,fill=black] (\from) -- (\to);
\foreach \from/\to in {31/3141,31/3142,32/3242,32/3243,33/3343,33/3344}
\draw [-,line width=0.4mm,fill=black] (\from) -- (\to);
\foreach \from/\to in {3141/41,3142/42,3242/42,3243/43,3343/43,3344/44}
\draw [->,line width=0.4mm,fill=black] (\from) -- (\to);
%%% ZBDT
\tikzstyle{every node} = [circle, fill=blue!10,draw]
\node (20) at (3,-3) {$B_{1,0}$};
\node (30) at (6,-3) {$B_{2,0}$};
\node (40) at (9,-3) {$100$};
\tikzstyle{every node} = [circle, fill=blue!40]
\node (1120) at (1.5,-1.5) {$x_{0}$};
\node (2030) at (4.5,-3) {$x_{0}$};
\node (3040) at (7.5,-3) {$x_{0}$};
%%%%%%%%% 
\node (2130) at (4,-1) {$x_{0}$};
\node (2031) at (4,-2) {$x_{0}$};
\node (3140) at (7,-1) {$x_{0}$};
\node (3041) at (7,-2) {$x_{0}$};
%%%%%%%%% 
\foreach \from/\to in {11/1120}
\draw [-,line width=0.4mm,fill=black] (\from) -- (\to);
\foreach \from/\to in {1120/20}
\draw [->,line width=0.4mm,fill=black] (\from) -- (\to);
\foreach \from/\to in {20/2030}
\draw [-,line width=0.4mm,fill=black] (\from) -- (\to);
\foreach \from/\to in {2030/30}
\draw [->,line width=0.4mm,fill=black] (\from) -- (\to);
\foreach \from/\to in {30/3040}
\draw [-,line width=0.4mm,fill=black] (\from) -- (\to);
\foreach \from/\to in {3040/40}
\draw [->,line width=0.4mm,fill=black] (\from) -- (\to);
%%%%%%%%% 
\foreach \from/\to in {20/2031,21/2130,30/3041,31/3140}
\draw [-,line width=0.4mm,fill=black] (\from) -- (\to);
\foreach \from/\to in {2130/30,2031/31,3140/40,3041/41}
\draw [->,line width=0.4mm,fill=black] (\from) -- (\to);
\end{tikzpicture}
%\caption{Prices $P_{i,j}$ in the ZBDT tree for three periods, and the corresponding transition probabilities
%for a zero-coupon bond with face value 100.}\label{figure:zbdtp}
}
\caption{The ZBDT interest rate tree (a) and the corresponding zc-bonds (b) with $T=3$ and $FV=100$.}\label{figure:zbdtr}
\end{figure}

Our modification of the classical BDT interest rate tree model 
adds to the dynamics the possibility of a downwards jump with a small probability at each time step
to a practically zero interest rate, 
where,
after its arrival, 
the process remains with high probability.
More precisely, in the new model, the nodes labeled $(i,j)$ with $j\geq 2$ have the same characteristics as in the BDT model 
(up and down probabilities $1/2$, and jump to values to be calibrated).
In addition, the nodes of the form $(1,j)$ add a third possible downwards jump with a small probability $p$
and the other two possible jumps have probability $\hat{p}=(1-p)/2$.
If this downwards jump is realized, then the process enters the so called ZIRP zone, 
meaning that  interest rate becomes a small value $x_0$ 
(close to the target of the policy). When the process is in the ZIRP zone, 
 it remains there with a high probability $(1-q)$ and exits with probability $q$.  Finally, to calibrate the tree,
following the same convention as in the classical BDT model,
we further impose that the variance at each node for the same time period remains the same 
(to be determined by calibration, denoted below by $\sigma(n)$ for the period $n$). 
To the previous $r_{i,j}$ and $B_{i,j}$ corresponding to the BDT model,
the ZBDT model adds the (unknown) bond prices $B_{i,0}$ for $i=1,\dots,n-1$ and $B_{0,n}=100$.
The corresponding interest rates $r_{i,0}$ for $1,\dots,i+1$ are fixed to  $x_0$.
In Figure \ref{figure:zbdtr} (b), we present the tree corresponding to the prices of a zc-bond 
with expiration in $n=3$ years. Note that $p$, $q$, $x_{0}$ have a clear interpretation as a probability of the crisis (in the basic period of time, which in this paper is $1$ year), a conditional probability of economic recovery from the financial crisis (in the basic period of time) and an assumed value of the interest rate in the ZIRP zone respectively.

\subsection{Calibration of the ZBDT model}

For the calibration, we use the same data as in the BDT model.
The strategy is modified to cope with the new unknowns, but follows the same
ideas. 
The first step uses only $y(1)$ and we conclude that $r_{01}=y(1)$.
The equations are
$$
B_{1,0}=B_{1,1}=B_{1,2}=100,
$$
$$
B_{0,1}=\frac{100}{1+y(1)}=\frac{1}{1+r_{0,1}}\left(\frac{1-p}2\left(B_{1,2}+B_{1,1}\right)+pB_{1,0}\right).
$$
 
\subsubsection*{Variance equation at a node}
In the present situation, the random variable $y$ takes three values:
$$
Y=
\begin{cases}
y_u,&\text{ with probability}\  \hat{p},\\
y_d, &\text{ with probability}\  \hat{p},\\
y_0, &\text{ with probability}\  p.\\
\end{cases}
$$
Then, $\log Y$ has the same variance as the random variable
$$
\log {Y\over y_0}=
\begin{cases}
\log {y_u \over y_0},&\text{ with probability}\   \hat{p},\\
\log {y_d \over y_0}, &\text{ with probability}\  \hat{p},\\
0, &\text{ with probability}\  p.\\
\end{cases}
$$
The mean of $\log(Y/y_0)$ is
$$
\frac{1-p}2\left(\log{y_u\over y_0}+\log{y_d\over y_0}\right),
$$
then
$$
\var\log {Y\over y_0} =\frac{1-p}2\left(\left(\log{y_u\over y_0}\right)^2+\left(\log{y_d\over y_0}\right)^2\right)-
\left(
\frac{1-p}2\left(\log{y_u\over y_0}+\log{y_d\over y_0}\right)
\right)^2.
$$
Introducing the notation
\begin{equation}\label{eq:ells}
\ell_u=\log\frac{y_{u}}{y_{0}},\quad
\ell_d=\log\frac{y_{d}}{y_{0}}.
\end{equation}
we obtain
\begin{equation}\label{eq:tres}
\var\log y=
\frac{1-p^2}4\left(\ell_u^2+\ell_d^2\right)-\frac{(1-p)^2}2\ell_u\ell_d.
\end{equation}
Considering now the interest rates, if the node $n-1,j$ has two edges (i.e. $j=2,\dots,n$), then
the variance at the node satisfies equation \eqref{eq:sigma},
(the same as in the BDT case). 
If the node has three edges (when $j=1$), then
the variance satisfies equation \eqref{eq:tres}.

%\subsubsection{The equations for general $n$}
%\subsubsection{Step $n=2$}
In the second step of calibration, the new data are $y(2)$ and $\beta(2)$.
The unknowns are $r_{1,1},r_{1,2}$ and $\sigma(2)$.
In this case 
$r_{1,1}=y_d$, 
$r_{1,2}=y_u$, $y_0=x_0$ and $\sigma(2)=\beta(2)$ 
(because in this case the local variation of the interest rate for one year coincides with the global variation).
Accordingly, $y_d=r_{1,1}$ and $y_u=r_{1,2}$.

\begin{align*}
B_{0,1}&=\frac{1}{{(1+y(2))}^2}=\frac{1}{1+r_{0,1}}\left(\frac{1-p}2\left(B_{u}+B_{d}\right)+pB_{0}\right),\\
B_{2,j}&=100, \quad j=1,2,3,\\
B_{u}&=\frac{100}{{(1+y_u)}},%\\
\quad
B_{d}=\frac{100}{{(1+y_d)}},%\\
\quad
B_{0}=\frac{100}{{(1+y_0)}},\\
\beta(2)^2&=\frac{1-p^2}4\left(\ell_u^2+\ell_d^2\right)-\frac{(1-p)^2}2\ell_u\ell_d.\\
\end{align*}
with $\ell_u$ and $\ell_d$ given in equation \eqref{eq:ells}.

For general $n$, the new data are $y(n)$ and $\beta(n)$. The unknowns are $r_{n-1,j}$ for $j=1,\dots,n$,
and $\sigma(n)$. The calibration equations are:
{\allowdisplaybreaks
\begin{align*}
B_{0,1}&=\frac{1}{{(1+y(n))}^n}=\frac{1}{1+r_{0,1}}\left(\frac{1-p}{2}B_{u}+\frac{1-p}{2}B_{d}+pB_{0}\right),\\
B_{n,j}&=100, \  j=0,\dots,(n+1),\\
B_{i,j}&=\frac12\frac{1}{1+r_{i,j}}\left(B_{i+1,j+1}+B_{i+1,j}\right),\  i=1,\dots,(n-1), \ j=2,\dots,i+1,\\
B_{i,1}&=\frac{1}{1+r_{i,j}}\left(\frac{1-p}{2}B_{i+1,2}+\frac{1-p}{2}B_{i+1,1}+pB_{i+1,0}\right),\  i=1,\dots,(n-1),\\
B_{i,0}&=\frac{1}{1+x_{0}}\left(qB_{i+1,1}+(1-q)B_{i+1,0}\right),\  i=1,\dots,(n-1),\\
B_{u}&=\frac{100}{{(1+y_u)}^{n-1}},
\quad
B_{d}=\frac{100}{{(1+y_d)}^{n-1}},
\quad
B_{0}=\frac{100}{{(1+y_0)}^{n-1}},\\
\ell_u&=\log\frac{y_{u}}{y_{0}},\quad
\ell_d=\log\frac{y_{d}}{y_{0}},\\
\beta(n)^2&=\frac{1-p^2}4\left(\ell_u^2+\ell_d^2\right)-\frac{(1-p)^2}2\ell_u\ell_d,\\
\sigma(n)&=\frac{1}{2}\log{\frac{r_{n-1,j+1}}{r_{n-1,j}}},\  j=2,\dots,n,\\
\ell_{1}&=\log\frac{r_{n-1,1}}{x_{0}},\qquad\ell_{2}=\log\frac{r_{n-1,2}}{x_{0}},\\
\sigma(n)^2&=\frac{1-p^2}4\left(\ell_1^2+\ell_2^2\right)-\frac{(1-p)^2}2\ell_1\ell_2.\\
\end{align*}}

%\newpage

\section{Empirical analysis of different scenarios with US treasury bonds data}\label{section:empirical}
The main motivation of our work is to analyze  
the new features observed in bond prices as a consequence 
of the ZIRP implemented
by the US Government in 2008.
In the Timeline \ref{timeline}, 
we give an account of the main events related with the US economy during 
the period of the study.
%\subsection{A timeline of financial events and corresponding yield-volatility curves}
\begin{table}%[H]
\caption{Timeline of relevant financial events 2000 - 2016.}\label{timeline}
\centering
\begin{minipage}[t]{.99\linewidth}
\color{gray}
\rule{\linewidth}{1pt}
\ytl{2000-2001}{Bursting of the dot.com and the Corporate Fraud}
\ytl{2002-2003}{US economy resumed expanding, while inflation rate and interest rate remained relatively low}
\ytl{2004-2006}{US economy expansion. The Federal Reserve hiked the interest rate in 17 consecutive times}
\ytl{2007}{Sub-prime housing crisis. Large financial institution were holding portfolios of loans that were worthless}
\ytl{2008}{US financial crisis. The Federal Reserve decreased the interest rate to 0-0.25\%}
\ytl{2009}{US economic recession}
\ytl{2010}{Exacerbation of the financial crisis in Europe}
\ytl{2011-2014}{Continues the policy of low interest rate. Medium volatilities rates}
\ytl{2015-2016}{Economic growth. The Federal Reserve increased its interest rate twice by 0.25\%}

\bigskip
\rule{\linewidth}{1pt}%
\end{minipage}%
\end{table}
%\newpage

\subsection{Interest rates yields and volatilities 2002-2017}

\begin{figure}[H]
\centering 
\includegraphics[width=8.4cm,height=6.0cm]{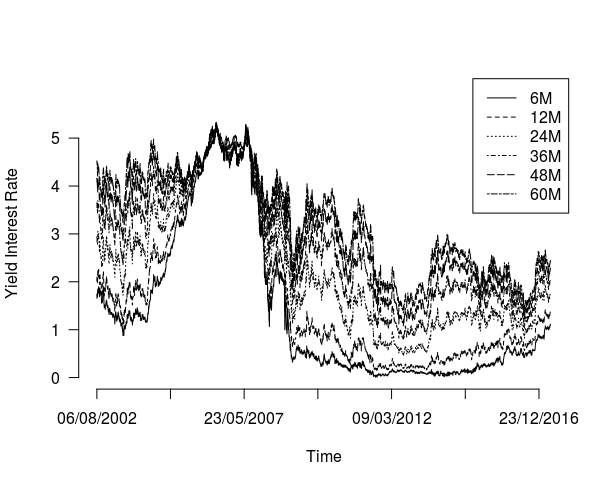}
\includegraphics[width=8.4cm,height=6.0cm]{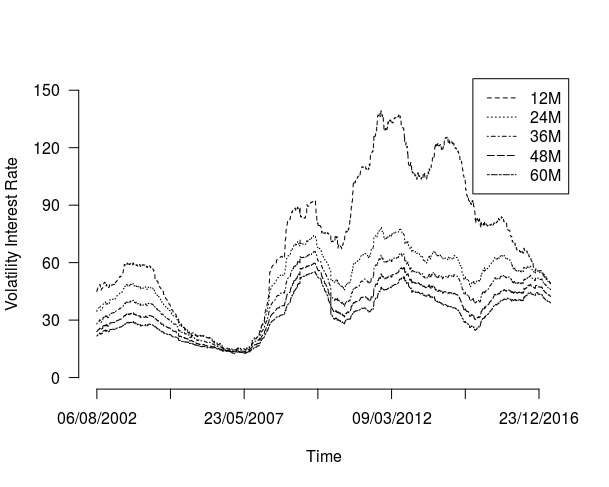}
\caption{Yield rates and yield volatilities for different maturities (2002-2017).}\label{figure:vol}
\end{figure}

We present the yields and its volatilities used to calibrate the ZBDT model (including interest rates and bond prices) 
with the aim of computing  bond option prices. 
The daily interest rates correspond to the period from August 6, 2002 to April 28, 2017,
and were obtained from the Federal Reserve Board of the United States.
The data are denoted by $y(t,k)$, where $t$ denotes the day and $k$ the corresponding six maturities used in this study ($k=1/2,1,2,3,4,5$ in years). 
In the previous sections, $t$ was omitted
because the analysis was performed for a fixed time.
To compute the volatility $\beta(t,k)$ corresponding to these values, we use the formulas 
$$
\aligned
\bar{\ell}(t,k)&=\frac1{252}\sum_{i=0}^{251}\log{y(t-i,k)\over y(t-i-1,k)},\\
\beta^2(t,k)&=\sum_{i=0}^{251}\left(\log{y(t-i,k)\over y(t-i-1,k)}-\bar{\ell}(t,k)\right)^2,
\endaligned
$$

where the factor 252 corresponds to the number of business day of one year;
that is, the window chosen to compute the volatilities.
The obtained data is presented in  Figure \ref{figure:vol}. 
%\textcolor{red}{different curves in figures can not be differ..Is it possible to change legend?}.

%\subsection{Scenarios}
%
%We analyze six different scenarios and explore how the ZBDT implied volatility values differ from the respective BDT values

 %\newpage 
\subsection{Six observed typical scenarios}

We choose six different days, corresponding to six different periods, expecting to 
analyze the impact of the downwards jump in the interest rate included in the ZBDT model.
To select each of these days, the interest rates depicted in the Figure \ref{figure:vol} and
the Timeline \ref{timeline} were taken into account.

In each of the six chosen scenarios, we calibrate the BDT and ZBDT models,
presenting the corresponding interest rates and bond prices.
These numerical results can be seen in Tables 
\ref{table:rp2}, \ref{table:rp4}, \ref{table:rp6}, \ref{table:rp8} \ref{table:rp10} and \ref{table:rp12} in the Appendix.

With this information, we compute vanilla call option prices along strikes of bond prices ranging from 80 to 100,
obtaining the respective implied volatility.  
To compute the implied volatility at time $t$ of an option written on a zc-bond that expires at time $T$, 
with strike $K$ and maturity $S$ ($t<S<T$), we use Black's formula (see (Black, 1976)),
which states
$$
C=B(t,T)\Phi(d_1)-KB(t,S)\Phi(d_2),
$$
where
$$
d_{1,2}={\log\left({B(t,T)\over KB(t,S)}\right)\over\sigma\sqrt{S-t}}\pm{\sigma\sqrt{S-t}\over 2},
$$
and  $\Phi$ is the  cumulative normal distribution function. Note that regardless of the model under consideration, we assume that the implied volatility is equal to $0$ if the corresponding option is worthless. For more details see  (McDonald, 2006).
In our empirical exercise, we consider a zc-bond with expiration in five years  ($T=5$) 
and European call options written at $t=0$ with exercise time two years ($S=2$). For the ZBDT model, we assume  the parameters $x_{0}=0.25\%$, $p=0.02$, $q=0.07$.
The results are presented in Tables \ref{table:rp3},  \ref{table:rp5}, \ref{table:rp7}, \ref{table:rp9}, \ref{table:rp11} and \ref{table:rp13} in the Appendix.
A primary conclusion is that, in contrast to the BDT model, the ZBDT allows us to price options with strikes
close to the face value of the bond, which corresponds to low interest rate periods.
This gives more accurate option prices in pre-crisis periods.

\section{Conclusions}\label{section:conclusions}

In the present work, we propose a novel and practical approach to model the 
possibility of a drop in the interest rates structure of sovereign bonds.
This modification is motivated by the recent 2008--2009 crisis in United States.

Our approach is inspired by  Lewis's (2016) ZIRP models   in  continuous time, 
and also in
Duffie and Singleton's (1999) default framework of bond pricing models. 
Our proposal consists in adding a new
branch at each period to the classical Black-Derman-Toy tree model that takes into account the small probability of this drop event to happen.
We name this the ZBDT model, the ``Z'' standing for (close to)  interest rate.
To the best of our knowledge, our model is the first discrete space--time model proposed for the ZIRP, 
and it shares 
the motivation of including this phenomena as previously considered in continuous time
models through sticky diffusions (such as in (Lewis, 2016)) and skew diffusions (Tian \&  Zhang, 2018).

This paper includes a development of the corresponding modified calibration scheme 
(that, naturally, happens to be more complex than the classical BDT calibration, and uses the same information) 
to obtain the interest rate tree and corresponding bond prices.
With this information, we valuate European option prices provided by both models.
The comparison between the two models is carried out though the implied volatility analysis provided by the Black option pricing formula. 
Our proposal opens the possibility of correcting option prices in different scenarios, 
especially under the risk of future  interest rates. 
The analysis of implied volatility curves provided by the US bond 
market is a tool that can reveal in which situation this drop probability is not negligible. 
Our main conclusion is that the ZIRP models allows u to price options with high strikes.
All of the observed implied volatilities are higher in the ZBDT model than in the BDT model.
This gives more accurate option prices in pre-crisis periods.

Further research includes the consideration of American bond options market prices (and possible other usual derivatives in
the bond markets) to calibrate the parameters of the proposed model: the probability of drop and  the probability of staying in the ZIRP zone,
and the more complex task of proposing a continuous time model analog.

    \bibliographystyle{agsm}
  % \bibliography{zbdt.25}

\begin{thebibliography}{99}
\addtolength{\leftmargin}{0.2in} % sets up alignment with the following line.
\setlength{\itemindent}{-0.2in}

\bibitem{black} Black, F. (1976), \emph{The Pricing of Commodity Contracts.} Journal of Financial Economics , Vol 3, pp 167-179.

\bibitem{bdt} Black, F.; Derman, E. \& Toy, W. (1990), \emph{A One-Factor Model of Interest Rates and Its Application to Treasury Bond Options} Financial Analysts Journal, Vol 46, No 1, pp 33-39.

\bibitem{borodin-salminen}
Borodin, A. N.;  Salminen, P. (2002),
\emph{Handbook of Brownian motion-facts and formulae.}
Second edition.
Probability and its Applications. Birkhäuser Verlag, Basel,  2002. 

\bibitem{bm}%useless
Brigo, D.,  Mercurio, F. (2006), 
 \emph{Interest Rate Models  Theory and Practice with Smile, Inflation and Credit.}
 Second Edition Springer Verlag.

\bibitem{cir}
Cox, J., Ingresoll J. \& Ross S. (1985),
\emph{A Theory of the Term Structure of Interest Rate.}
Econometrica 53, pp. 385-407.

\bibitem{DuffieSingleton}
Duffie, D., Singleton, K. (1999),
\emph{Modeling Term Structure of Defaultable Bonds.}
Review of Financial Studies, Vol 12, pp  687-720.

\bibitem{eberlein}
Eberlein, E., Gerhart, G. \& Grbac, Z. (2018),
\emph{Multiple Curve L\'evy Forward Price Model Allowing for Negative Interest Rates.}
Quantitative Finance, Vol 18,  Issue 4.

\bibitem{fili}%useless?
Filipovic, D. (2009), 
\emph{Term Structure Models.}
Springer Finance.

\bibitem{hull}%useless?
 Hull, J. (2009)
\emph{Technical Note No. 23. Options, Futures, and  Other Derivatives.}
Options, Futures, and Other Derivatives.

\bibitem{lejay}
Lejay, A. (2006),
\emph{On the Constructions of the Skew Brownian Motion.}
Probability Surveys, Vol 3, pp 413-466.

\bibitem{Lewis}
Lewis, A. (2016),
\emph{Option Valuation under Stochastic Volatility II.}
 Finance Press, Newport Beach, California, USA.

\bibitem{martin}
Martin, M. (2018),
\emph{An Overview of Post-crisis Term Structure Models.}
New Methods in Fixed Income Modeling, Springer, pp 85-97.

\bibitem{mcd}
 McDonald, R. (2006),
\emph{Derivatives Markets.} 
Third edition, Pearson Series in Finance, Boston: Addison-Wesley, 2006

\bibitem{skew}
Tian, Y., Zhang, H. (2018),
\emph{Skew CIR Process, Conditional Characteristic Function, Moments and Bond Pricing.}
Applied Mathematics and Computation, Vol 329, pp 230-238.

\end{thebibliography}
   %\nocite{*}

\newpage
\section{Appendix}
\subsubsection*{Scenario I (May 23, 2003): Expanding economy, normal term structure.}  
\begin{table}[h]
\centering
\scalebox{0.75}{
\begin{tabular}{|ccccc|cccccc|}
\hline
	&          &         &          &             &{}     &              &       &       &  &100\\ 
          &        &         &          &9.32 	&  {} &              &       & &91.47  &100\\
          &        &         & 8.34 & 6.56 	&   {}    &              & &85.52 &93.84  &100\\
          &         & 6.52 & 5.30 & 4.62	& {}       &        &82.36 &89.95 &95.58  &100\\
          & 3.76 &3.50 & 3.36 & 3.26	&  {}    & 82.30 &88.42 &93.08 &96.85  &100\\     
 1.36 & 1.54 & 1.87& 2.13 & 2.29	&{} 84.53& 89.05& 92.44 &95.27 &97.76  &100\\
 	\hline    \hline
	&          &         &          &             &{}           &   &       &       &  &100\\
       &         &          &             & 21.24   &{}          &    &       & &82.48  &100\\
        &         &          &  14.57 & 8.81        &{}           &   & &76.10 &91.90  &100\\
                &         & 8.91 &  5.51 & 3.65    &{}     &   &75.92 &89.27 &96.48  &100\\
                    & 4.21 & 2.92 &  2.09 &  1.52       &{} &79.49 &89.76 &95.50 &98.51  &100\\
                    1.36 & 1.13 & 0.97  & 0.80 & 0.63    &{}84.53& 91.81 &95.90&98.16 &99.37  &100\\
                    & 0.25 & 0.25 & 0.25 &0.25   		&{}  & 98.76 &99.17 &99.48 &99.75  &100\\
\hline
\end{tabular}
}
\caption{ BDT interest rates (top-left), BDT bond prices (top-right), ZBDT interest rates (bottom-left) and  ZBDT bond prices (bottom-right) in scenario I.}\label{table:rp2}
\end{table}

%%%%%%%%%%%%%%%%
\begin{table}[!ht]\footnotesize
\begin{center}
\scalebox{0.85}{
\begin{tabular}{|c|c|c|c|c|}
\hline
 strikes  	&	 BDT    	&	$\textbf{v}$	&	ZBDT	&	$\textbf{v}$	\\
\hline
80	&	7.6328	&	1.5372	&	8.5965	&	1.5691	\\
81	&	6.6716	&	1.5058	&	7.8715	&	1.5479	\\
82	&	5.7103	&	1.4714	&	7.1465	&	1.5254	\\
83	&	4.9003	&	1.4399	&	6.4214	&	1.5014	\\
84	&	4.1768	&	1.409	&	5.6964	&	1.4757	\\
85	&	3.4533	&	1.3744	&	4.9714	&	1.4477	\\
86	&	2.7298	&	1.3343	&	4.2463	&	1.4170	\\
87	&	2.0063	&	1.2862	&	3.5213	&	1.3826	\\
88	&	1.2827	&	1.2236	&	2.7963	&	1.3430	\\
89	&	0.8363	&	1.1714	&	2.0713	&	1.2955	\\
90	&	0.5934	&	1.1343	&	1.4609	&	1.2458	\\
91	&	0.3505	&	1.0825	&	1.2150	&	1.2226	\\
92	&	0.1076	&	0.9855	&	0.9691	&	1.1955	\\
93	&	0.00	&	0.00	&	0.7235	&	1.1624	\\
94	&	0.00	&	0.00	&	0.4774	&	1.119	\\
95	&	0.00	&	0.00	&	0.2315	&	1.0522	\\
96	&	0.00	&	0.00	&	0.009	&	0.8459	\\
97	&	0.00	&	0.00	&	0.0062	&	0.8300	\\
98	&	0.00	&	0.00	&	0.0033	&	0.8049	\\
99	&	0.00	&	0.00	&	0.0005	&	0.7364	\\
\hline
\end{tabular}}
\caption{Call option prices and implied volatility  for both models in scenario I.}\label{table:rp3}
\end{center}
\end{table}
%\end{minipage}
%\hfill
%\begin{minipage}[b]{0.55\linewidth}
%\centering
%\begin{figure}[H]
%\includegraphics[scale=0.38]{imp_vol_1}
%\caption{Implied volatility for both models in scenario I
%(BDT-red, ZBDT-black).}\label{figure:iv1}
%\end{figure}
%\end{minipage}
%\end{table}
\newpage
\subsubsection*{Scenario II (August 07, 2006): Flat term structure curves. }

\begin{table}[h]
\centering
\scalebox{0.8}{
\begin{tabular}{|ccccc|cccccc|}
\hline
		& 		&    	&       & 		&{}     &		& 		&  		&		&100\\ 
        &      	&       &       & 8.41	&{} 	&     	&   	& 		&92.24  &100\\
        &     	&       & 7.12	& 6.40	&{}    	&    	& 		&86.92 	&93.98  &100\\
        &      	& 6.37	& 5.51	& 4.88	&{}     &       &83.04 	&89.72 	&95.35  &100\\
        & 5.66	& 4.77	& 4.26	& 3.71	&{}    	& 80.33 &86.71 	&91.96 	&96.42  &100\\     
4.97	& 3.96	& 3.57	& 3.30	& 2.83	& 78.66 & 84.82 & 89.65 & 93.74 &97.25 	& 100\\
 	\hline    \hline
		& 		&    	&       & 		&{}     &		& 		&  		&		&100\\ 
        &      	&       &       & 17.79	&{} 	&     	&   	& 		&84.89  &100\\
        &     	&       & 12.23	& 8.58	&{}    	&    	& 		&78.86 	&92.09  &100\\
        &      	& 9.06	& 6.08	& 4.14	&{}     &       &76.80 	&88.67 	&96.02  &100\\
        & 6.70	& 4.32	& 3.02	& 2.00	&{}    	& 77.06 &87.65 	&94.19 	&98.04  &100\\     
4.97	& 3.00	& 2.13	& 1.54	& 0.98	& 78.66 & 88.02 &93.63	&97.04	&99.03 	&100\\
		& 0.25	& 0.25	& 0.25	& 0.25	& 		& 98.58 &99.10	&99.47	&99.75 	&100\\
\hline
\end{tabular}}
\caption{BDT interest rates (top-left), BDT bond prices (top-right), ZBDT interest rates (bottom-left) and  ZBDT bond prices (bottom-right) in scenario II.}\label{table:rp4}
\end{table}

%%%%%%%%%%%

%\begin{table}[h]
%\begin{minipage}[b]{0.45\linewidth}
%\centering
%\scalebox{0.55}{
\begin{table}[!ht]\footnotesize
\begin{center}
\scalebox{0.94}{
\begin{tabular}{|c|c|c|c|c|}
\hline
 strikes  	&	 BDT    	&	$\textbf{v}$	&	ZBDT	&	$\textbf{v}$	\\
\hline
80	&	5.9450	&	1.4960	&	6.6568	&	1.5239	\\
81	&	5.0361	&	1.4600	&	5.9706	&	1.4995	\\
82	&	4.1271	&	1.4196	&	5.2844	&	1.4734	\\
83	&	3.2181	&	1.3731	&	4.5982	&	1.4449	\\
84	&	2.5267	&	1.3320	&	3.9120	&	1.4135	\\
85	&	1.8431	&	1.2828	&	3.2258	&	1.3782	\\
86	&	1.1595	&	1.2184	&	2.5396	&	1.3374	\\
87	&	0.6079	&	1.1417	&	1.8534	&	1.2881	\\
88	&	0.3788	&	1.0940	&	1.3277	&	1.2412	\\
89	&	0.1497	&	1.0135	&	1.0945	&	1.2171	\\
90	&	0.00		&	0.00		&	0.8614	&	1.1887	\\
91	&	0.00		&	0.00		&	0.6283	&	1.1535	\\
92	&	0.00		&	0.00		&	0.3951	&	1.1061	\\
93	&	0.00		&	0.00		&	0.1620	&	1.0272	\\
94	&	0.00		&	0.00		&	0.0139	&	0.8702	\\
95	&	0.00		&	0.00		&	0.0111	&	0.8607	\\
96	&	0.00		&	0.00		&	0.0085	&	0.8487	\\
97	&	0.00		&	0.00		&	0.0057	&	0.8322	\\
98	&	0.00		&	0.00		&	0.0030	&	0.8059	\\
99	&	0.00		&	0.00		&	0.0003	&	0.7224	\\
\hline
\end{tabular}}
\caption{Call option prices and implied volatility  for both models in scenario II.}\label{table:rp5}
\end{center}
\end{table}
%\end{minipage}
%\hfill
%\begin{minipage}[b]{0.45\linewidth}
%\centering
%\begin{figure}[H]
%\includegraphics[scale=0.33]{imp_vol_2}
%\caption{Implied volatility for both models in scenario II.}\label{figure:iv2}
%\end{figure}
%\end{minipage}
%\end{table}

\newpage 
\subsubsection*{Scenario III (November 14, 2007): Start of financial crisis.}      

\begin{table}[h]
\centering
\scalebox{0.8}{
\begin{tabular}{|ccccc|cccccc|}
\hline
		& 		&    	&       & 		&{}     &		& 		&  		&		&100\\ 
        &      	&       &       & 7.91	&{} 	&     	&   	& 		&92.67  &100\\
        &     	&       & 7.03	& 6.08	&{}    	&    	& 		&87.33 	&94.27  &100\\
        &      	& 6.06	& 5.24	& 4.68	&{}     &       &83.68 	&90.17 	&95.53  &100\\
        & 4.73	& 4.23	& 3.91	& 3.60	&{}    	& 81.77 &87.59 	&92.42 	&96.53  &100\\     
3.56	& 3.06	& 2.95	& 2.91	& 2.77	& 81.22 & 86.46 &90.62  &94.17  &97.31 	&100\\
 	\hline    \hline
		& 		&    	&       & 		&{}     &		& 		&  		&		&100\\ 
        &      	&       &       & 16.86	&{} 	&     	&   	& 		&85.57  &100\\
        &     	&       & 12.18	& 8.18	&{}    	&    	& 		&79.34 	&92.44  &100\\
        &      	& 8.62	& 5.72	& 3.97	&{}     &       &77.59 	&89.21 	&96.19  &100\\
        & 5.58	& 3.75	& 2.69	& 1.92	&{}    	& 78.69 &88.58 	&94.60 	&98.11  &100\\     
3.56	& 2.26	& 1.68	& 1.29	& 0.95	& 81.22 & 89.48 &94.39	&97.33	&99.06 	&100\\
		& 0.25	& 0.25	& 0.25	& 0.25	& 		& 98.63 &99.11	&99.47	&99.75 	&100\\
\hline
\end{tabular}}
%}
\caption{BDT interest rates (top-left), BDT bond prices (top-right), ZBDT interest rates (bottom-left) and  ZBDT bond prices (bottom-right) in scenario III.}\label{table:rp6}
\end{table}

%%%%%

%\begin{table}[h]
%\begin{minipage}[b]{0.45\linewidth}
%\centering
%\scalebox{0.55}{
\begin{table}[!ht]\footnotesize
\begin{center}
\scalebox{0.94}{
\begin{tabular}{|c|c|c|c|c|}
\hline
 strikes  	&	 BDT    	&	$\textbf{v}$	&	ZBDT	&	$\textbf{v}$	\\
\hline
80	&	6.8635	&	1.5215	&	7.4143	&	1.5414	\\
81	&	5.9340	&	1.4884	&	6.7131	&	1.5185	\\
82	&	5.0046	&	1.4519	&	6.0118	&	1.4940	\\
83	&	4.0751	&	1.4110	&	5.3105	&	1.4677	\\
84	&	3.2198	&	1.3676	&	4.6092	&	1.4390	\\
85	&	2.5208	&	1.3265	&	3.9079	&	1.4073	\\
86	&	1.8218	&	1.2765	&	3.2066	&	1.3715	\\
87	&	1.1229	&	1.2103	&	2.5053	&	1.3300	\\
88	&	0.6141	&	1.1396	&	1.8041	&	1.2794	\\
89	&	0.3799	&	1.0014	&	1.2969	&	1.2339	\\
90	&	0.1456	&	1.0090	&	1.0589	&	1.2089	\\
91	&	0.00		&	0.00		&	0.8209	&	1.1792	\\
92	&	0.00		&	0.00		&	0.5829	&	1.1419	\\
93	&	0.00		&	0.00		&	0.3449	&	1.0897	\\
94	&	0.00		&	0.00		&	0.1069	&	0.9926	\\
95	&	0.00		&	0.00		&	0.0114	&	0.8587	\\
96	&	0.00		&	0.00		&	0.0086	&	0.8467	\\
97	&	0.00		&	0.00		&	0.0059	&	0.8304	\\
98	&	0.00		&	0.00		&	0.0031	&	0.8044	\\
99	&	0.00		&	0.00		&	0.0003	&	0.7252	\\
\hline
\end{tabular}}
\caption{Call option prices and implied volatility  for both models in scenario III.}\label{table:rp7}
\end{center}
\end{table}
%\end{minipage}
%\hfill
%\begin{minipage}[b]{0.45\linewidth}
%\centering
%\begin{figure}[H]
%\includegraphics[scale=0.33]{imp_vol_3}
%\caption{Implied volatility for both models in scenario III.}\label{figure:iv3}
%\end{figure}
%\end{minipage}
%\end{table}

\newpage

\subsubsection*{Scenario IV (August 08, 2008): US crisis.}

\begin{table}[h]
\centering
\scalebox{0.8}{
\begin{tabular}{|ccccc|cccccc|}
\hline
		& 		&    	&       & 		&{}     &		& 		&  		&		&100\\ 
        &      	&       &       & 10.87	&{} 	&     	&   	& 		&90.20  &100\\
        &     	&       & 9.70	& 7.10	&{}    	&    	& 		&83.67 	&93.37  &100\\
        &      	& 8.04	& 5.67	& 4.64	&{}     &       &80.10 	&89.40 	&95.57  &100\\
        & 5.43	& 3.84	& 3.32	& 3.03	&{}    	& 79.69 &87.93 	&93.22 	&97.06  &100\\     
2.47	& 1.86	& 1.85	& 1.94	& 1.98	& 82.16 & 88.70 &92.76  &95.70  &98.06 	&100\\
 	\hline    \hline
		& 		&    	&       & 		&{}     &		& 		&  		&		&100\\ 
        &      	&       &       & 25.29	&{} 	&     	&   	& 		&79.82  &100\\
        &     	&       & 16.84	& 9.42	&{}    	&    	& 		&73.27 	&91.39  &100\\
        &      	& 10.72	& 5.80	& 3.51	&{}     &       &73.21 	&88.85 	&96.61  &100\\
        & 5.95	& 3.17	& 2.00	& 1.39	&{}    	& 76.77 &89.46 	&95.75 	&98.71  &100\\     
2.47	& 1.40	& 0.95	& 0.69	& 0.49	& 82.16 & 91.56 &96.18	&98.43	&99.51 	&100\\
		& 0.25	& 0.25	& 0.25	& 0.25	& 		& 98.79 &99.19	&99.49	&99.75 	&100\\
\hline
\end{tabular}}
%}
\caption{BDT interest rates (top-left), BDT bond prices (top-right), ZBDT interest rates (bottom-left) and  ZBDT bond prices (bottom-right) in scenario IV.}\label{table:rp8}
\end{table}

%\begin{table}[h]
%\begin{minipage}[b]{0.45\linewidth}
%\centering
%\scalebox{0.55}{
\begin{table}[!ht]\footnotesize
\begin{center}
\scalebox{0.94}{
\begin{tabular}{|c|c|c|c|c|}
\hline
 strikes  	&	 BDT    	&	$\textbf{v}$	&	ZBDT	&	$\textbf{v}$	\\
\hline
80	&	6.8145	&	1.5171	&	8.3751	&	1.5714	\\
81	&	6.0812	&	1.4919	&	7.6631	&	1.5500	\\
82	&	5.3708	&	1.4655	&	6.9511	&	1.5272	\\
83	&	4.6603	&	1.4369	&	6.2391	&	1.5030	\\
84	&	3.9499	&	1.4051	&	5.5271	&	1.4769	\\
85	&	3.2395	&	1.3693	&	4.8151	&	1.4486	\\
86	&	2.5291	&	1.3277	&	4.1031	&	1.4175	\\
87	&	1.8187	&	1.2770	&	3.3911	&	1.3825	\\
88	&	1.1398	&	1.2132	&	2.6791	&	1.3421	\\
89	&	0.9003	&	1.1853	&	1.9670	&	1.2935	\\
90	&	0.6608	&	1.1508	&	1.5078	&	1.2558	\\
91	&	0.4213	&	1.1048	&	1.2653	&	1.2334	\\
92	&	0.1818	&	1.0299	&	1.0227	&	1.2074	\\
93	&	0.00		&	0.00		&	0.7802	&	1.1760	\\
94	&	0.00		&	0.00		&	0.5376	&	1.1359	\\
95	&	0.00		&	0.00		&	0.2951	&	1.0780	\\
96	&	0.00		&	0.00		&	0.0525	&	0.9463	\\
97	&	0.00		&	0.00		&	0.0062	&	0.8323	\\
98	&	0.00		&	0.00		&	0.0033	&	0.8075	\\
99	&	0.00		&	0.00		&	0.0005	&	0.7417	\\
\hline
\end{tabular}}
\caption{Call option prices and implied volatility  for both models in scenario IV.}\label{table:rp9}
\end{center}
\end{table}
%\end{minipage}
%\hfill
%\begin{minipage}[b]{0.45\linewidth}
%\centering
%\begin{figure}[H]
%\includegraphics[scale=0.33]{imp_vol_4}
%\caption{Implied volatility for both models in scenario IV.}\label{figure:iv4}
%\end{figure}
%\end{minipage}
%\end{table}

\newpage
\subsubsection*{Scenario V (August 03, 2010): European crisis. } 
\begin{table}[h]
\centering
\scalebox{0.8}{
\begin{tabular}{|ccccc|cccccc|}
\hline
		& 		&    	&       & 		&{}     &		& 		&  		&		&100\\ 
        &      	&       &       & 9.65	&{} 	&     	&   	& 		&91.20  &100\\
        &     	&       & 8.04	& 6.33	&{}    	&    	& 		&85.73 	&94.05  &100\\
        &      	& 5.90	& 4.84	& 4.15	&{}     &       &83.27 	&90.65 	&96.01  &100\\
        & 2.89	& 2.98	& 2.91	& 2.72	&{}    	& 84.02 &89.63 	&93.95 	&97.35  &100\\     
0.51	& 1.13	& 1.50	& 1.75	& 1.79	& 86.87 & 90.60 &93.62  &96.12  &98.24 	&100\\
 	\hline    \hline
		& 		&    	&       & 		&{}     &		& 		&  		&		&100\\ 
        &      	&       &       & 21.45	&{} 	&     	&   	& 		&82.34  &100\\
        &     	&       & 13.76	& 8.22	&{}    	&    	& 		&76.80 	&92.40  &100\\
        &      	& 7.92	& 4.93	& 3.15	&{}     &       &77.38 	&90.22 	&96.94  &100\\
        & 3.27	& 2.46	& 1.77	& 1.21	&{}    	& 81.55 &90.96 	&96.17 	&98.81  &100\\     
0.51	& 0.84	& 0.77	& 0.64	& 0.47	& 86.87 & 93.03 &96.62	&98.54	&99.54 	&100\\
		& 0.25	& 0.25	& 0.25	& 0.25	& 		& 98.82 &99.20	&99.49	&99.75 	&100\\
\hline
\end{tabular}}
%}
\caption{BDT interest rates (top-left), BDT bond prices (top-right), ZBDT interest rates (bottom-left) and  ZBDT bond prices (bottom-right) in scenario V.}\label{table:rp10}
\end{table}

%\begin{table}[h]
%\begin{minipage}[b]{0.45\linewidth}
%\centering
%\scalebox{0.55}{
\begin{table}[!ht]\footnotesize
\begin{center}
\scalebox{0.94}{
\begin{tabular}{|c|c|c|c|c|}
\hline
 strikes  	&	 BDT    	&	$\textbf{v}$	&	ZBDT	&	$\textbf{v}$	\\
\hline
80	&	8.8374	&	1.5667	&	9.4675	&	1.5859	\\
81	&	7.8620	&	1.5378	&	8.7327	&	1.5659	\\
82	&	6.8866	&	1.5067	&	7.9978	&	1.5448	\\
83	&	5.9112	&	1.4730	&	7.2629	&	1.5225	\\
84	&	5.1113	&	1.4430	&	6.5280	&	1.4987	\\
85	&	4.3777	&	1.4128	&	5.7931	&	1.4731	\\
86	&	3.6440	&	1.3791	&	5.0582	&	1.4454	\\
87	&	2.9104	&	1.3405	&	4.3233	&	1.4149	\\
88	&	2.1767	&	1.2945	&	3.5884	&	1.3808	\\
89	&	1.4431	&	1.2358	&	2.8535	&	1.3415	\\
90	&	0.8912	&	1.1755	&	2.1186	&	1.2945	\\
91	&	0.6453	&	1.1402	&	1.4042	&	1.2362	\\
92	&	0.3993	&	1.0922	&	1.1555	&	1.2121	\\
93	&	0.1533	&	1.0101	&	0.9068	&	1.1836	\\
94	&	0.00		&	0.00		&	0.6581	&	1.1483	\\
95	&	0.00		&	0.00		&	0.4095	&	1.1003	\\
96	&	0.00		&	0.00		&	0.1608	&	1.0191	\\
97	&	0.00		&	0.00		&	0.0063	&	0.8279	\\
98	&	0.00		&	0.00		&	0.0034	&	0.8033	\\
99	&	0.00		&	0.00		&	0.0006	&	0.7389	\\
\hline
\end{tabular}}
\caption{Call option prices and implied volatility  for both models in scenario V.}\label{table:rp11}
\end{center}
\end{table}
%\end{minipage}
%\hfill
%\begin{minipage}[b]{0.45\linewidth}
%\centering
%\begin{figure}[H]
%\includegraphics[scale=0.33]{imp_vol_5}
%\caption{Implied volatility for both models in scenario V.}\label{figure:iv5}
%\end{figure}
%\end{minipage}
%\end{table}

 \newpage
 
\subsubsection*{Scenario VI (May 20, 2015): End of US-crisis.} 
\begin{table}[h]
\centering
\scalebox{0.8}{
\begin{tabular}{|ccccc|cccccc|}
\hline
		& 		&    	&       & 		&{}     &		& 		&  		&		&100\\ 
        &      	&       &       & 7.63	&{} 	&     	&   	& 		&92.91  &100\\
        &     	&       & 6.52	& 4.78	&{}    	&    	& 		&88.41 	&95.44  &100\\
        &      	& 5.34	& 3.72	& 2.99	&{}     &       &86.02 	&92.81 	&97.10  &100\\
        & 2.97	& 2.35	& 2.13	& 1.87	&{}    	& 86.46 &92.04 	&95.60 	&98.17  &100\\     
0.61	& 0.92	& 1.04	& 1.21	& 1.17	& 89.14 & 92.90 &95.47  &97.32  &98.84 	&100\\
 	\hline    \hline
		& 		&    	&       & 		&{}     &		& 		&  		&		&100\\ 
        &      	&       &       & 16.66	&{} 	&     	&   	& 		&85.72  &100\\
        &     	&       & 11.02	& 6.01	&{}    	&    	& 		&81.09 	&94.33  &100\\
        &      	& 6.91	& 3.70	& 2.17	&{}     &       &81.26 	&92.67 	&97.88  &100\\
        & 3.21	& 1.91	& 1.24	& 0.78	&{}    	& 84.53 &93.22 	&97.34 	&99.22  &100\\     
0.61	& 0.70	& 0.53	& 0.42	& 0.28	& 89.14 & 94.80 &97.68	&99.06	&99.72 	&100\\
		& 0.25	& 0.25	& 0.25	& 0.25	& 		& 98.91 &99.23	&99.50	&99.75 	&100\\
\hline
\end{tabular}}
%}
\caption{BDT interest rates (top-left), BDT bond prices (top-right), ZBDT interest rates (bottom-left) and  ZBDT bond prices (bottom-right) in scenario VI.}\label{table:rp12}
\end{table}

%\begin{table}[h]
%\begin{minipage}[b]{0.45\linewidth}
%\centering
%\scalebox{0.55}{
\begin{table}[!ht]\footnotesize
\begin{center}
\scalebox{0.94}{
\begin{tabular}{|c|c|c|c|c|}
\hline
 strikes  	&	 BDT    	&	$\textbf{v}$	&	ZBDT	&	$\textbf{v}$	\\
\hline
80	&	11.1312	&	1.6207	&	11.1312	&	1.6207	\\
81	&	10.1562	&	1.5959	&	10.1562	&	1.5959	\\
82	&	9.1811	&	1.5697	&	9.3581	&	1.5750	\\
83	&	8.2061	&	1.5419	&	8.6233	&	1.5551	\\
84	&	7.2310	&	1.5120	&	7.8886	&	1.5341	\\
85	&	6.2560	&	1.4796	&	7.1538	&	1.5118	\\
86	&	5.2809	&	1.4439	&	6.4190	&	1.4880	\\
87	&	4.5431	&	1.4146	&	5.6842	&	1.4625	\\
88	&	3.8094	&	1.3821	&	4.9494	&	1.4347	\\
89	&	3.0756	&	1.3450	&	4.2417	&	1.4040	\\
90	&	2.3419	&	1.3013	&	3.4799	&	1.3695	\\
91	&	1.6081	&	1.2467	&	2.7451	&	1.3297	\\
92	&	0.8744	&	1.1694	&	2.0103	&	1.2818	\\
93	&	0.6085	&	1.1302	&	1.2756	&	1.2188	\\
94	&	0.3623	&	1.0795	&	0.9197	&	1.1793	\\
95	&	0.1161	&	0.9860	&	0.6710	&	1.1447	\\
96	&	0.00		&	0.00		&	0.4222	&	1.0981	\\
97	&	0.00		&	0.00		&	0.1734	&	1.0205	\\
98	&	0.00		&	0.00		&	0.0035	&	0.7996	\\
99	&	0.00		&	0.00		&	0.0007	&	0.7397	\\
\hline
\end{tabular}}
\caption{Call option prices and implied volatility  for both models in scenario VI.}\label{table:rp13}
\end{center}
\end{table}

\end{document}